\documentclass[aps,prl,reprint,10pt,twocolumn,citeautoscript,superscriptaddress]{revtex4-1}
\pdfoutput=1

\usepackage{natbib}
\usepackage[english]{babel}
\usepackage{letltxmacro}
\usepackage{latexsym}
\LetLtxMacro{\ORIGselectlanguage}{\selectlanguage}
\makeatletter
\DeclareRobustCommand{\selectlanguage}[1]{%
  \@ifundefined{alias@\string#1}
    {\ORIGselectlanguage{#1}}
    {\begingroup\edef\x{\endgroup
       \noexpand\ORIGselectlanguage{\@nameuse{alias@#1}}}\x}%
}
\newcommand{\definelanguagealias}[2]{%
  \@namedef{alias@#1}{#2}%
}
\makeatother
\definelanguagealias{en}{english}
\definelanguagealias{English}{english}
\usepackage{graphicx}
\usepackage{amsmath}
\usepackage{amsfonts}
\usepackage{amssymb}
\usepackage{bm}
\usepackage{color}
\usepackage[percent]{overpic}
\usepackage{soul} 
\usepackage{amssymb}
\usepackage{wasysym}
\usepackage{dsfont}
\usepackage{float}
\usepackage[mathscr]{euscript}
\usepackage{lipsum}
\usepackage{hyperref}
\hypersetup{
    pdfstartview={FitH},    
    colorlinks=true,       
    linkcolor=blue,          
    citecolor=blue,        
    filecolor=magenta,      
    urlcolor=blue           
}

\newcommand{\pdagger}{\phantom{\dagger}}
\newcommand{\be}{\begin{equation}}
\newcommand{\ee}{\end{equation}}
\newcommand{\bea}{\begin{eqnarray}}
\newcommand{\eea}{\end{eqnarray}}

\newcommand{\vect}[1]{\boldsymbol{#1}}

\usepackage{pdfpages}
\usepackage{pgffor}
\makeatletter
\AtBeginDocument{\let\LS@rot\@undefined}
\makeatother

\usepackage{graphicx}
\usepackage[colorinlistoftodos]{todonotes}
\usepackage{verbatim}
\usepackage[normalem]{ulem}

\begin{document}

\title{Nagaoka ferromagnetism in doped Hubbard models in optical lattices}

\author{Rhine Samajdar}
\affiliation{Department of Physics, Princeton University, Princeton, NJ 08544, USA}
\affiliation{Princeton Center for Theoretical Science, Princeton University, Princeton, NJ 08544, USA}

\author{R. N. Bhatt}
\affiliation{Department of Physics, Princeton University, Princeton, NJ 08544, USA}
\affiliation{Department of Electrical and Computer Engineering, Princeton University, Princeton, NJ 08544, USA}

\date{\today}

\begin{abstract}
   The search for ferromagnetism in the Hubbard model has been a problem of outstanding interest since Nagaoka’s original proposal in 1966. Recent advances in quantum simulation have today enabled the study of tunable doped Hubbard models in ultracold atomic systems.  Employing large-scale density-matrix renormalization group calculations, we establish the existence of high-spin ground states of the Hubbard model on finite-sized triangular lattices, analyze the microscopic mechanisms behind their origin, and investigate the interplay between ferromagnetism and other competing orders, such as stripes. These results explain---and shed new light on---the intriguing observations of ferromagnetic correlations in recent optical-lattice experiments. Additionally, we examine a generalized variant of the Hubbard model, wherein any second electron on a single lattice site is weakly bound compared to the first one, and demonstrate how this modification can lead to enhanced ferromagnetism, at intermediate length scales, on the nonfrustrated square lattice as well. 
\end{abstract}

\maketitle

\textit{Introduction.---}The Hubbard model is a paradigmatic model of modern condensed matter physics that is widely used to study a variety of strongly correlated systems~\cite{arovas2022hubbard}. It describes electrons hopping on a lattice with a tunneling amplitude $t$ and interacting via an onsite potential $U$. Despite its seeming simplicity, this model harbors tremendously rich physics ~\cite{schafer2021tracking, qin2022hubbard} and is often believed to underlie a host of materials, including the high-temperature cuprate superconductors~\cite{lee2006doping}.

Over the last decade, ultracold atoms trapped in optical lattices have emerged as an exceptionally promising and versatile platform for realizing Hubbard models~\cite{esslinger2010fermi,tarruell2018quantum}. Recently, a number of such cold-atom experiments~\cite{cheuk2016observation,parsons2016site,drewes2017antiferromagnetic,mazurenko2017cold,spar2022realization} have demonstrated the existence of long-range antiferromagnetic order in two dimensions. Antiferromagnetism arises quite commonly in the Hubbard model, the magnetic excitations of which are described by a Heisenberg model  of spins at half filling (one electron per site)~\cite{manousakis1991spin} or, more generally, by a so-called $t$-$J$ model~\cite{chao1978canonical} in the doped system. In both cases, the  electron-electron interactions give rise to a superexchange $J$\,$\sim$\,$-t^2/U$~\cite{anderson1959new}, which favors antialignment of spins.

Away from half filling, the kinetic term of the model can actually favor \textit{ferromagnetism}, which is much rarer in the phase diagram. 
In 1966,~\citet{nagaoka1966ferromagnetism} rigorously proved that  for $U/t$\,$=$\,$\infty$, the ground state of the Hubbard model on a bipartite lattice with a single hole away from half filling is ferromagnetic~\cite{tian1990simplified}. For \textit{finite} $U/t$, however, in the thermodynamic limit, the ground state is antiferromagnetic at half filling in all integer dimensions $d$\,$>$\,$1$~\cite{sutHo1991absence}. Since then, a number of studies have sought to determine if and when ``Nagaoka ferromagnetism” is obtained for finite $U/t$~\cite{hanisch1993ferromagnetism,hanisch1995ferromagnetism,wurth1996ferromagnetism} or practically relevant dopings~\cite{PhysRevB.40.2719,fang1989holes, shastry1990instability,  basile1990stability,PhysRevB.41.11697} with varying degrees of success.  In the absence of a conclusive answer, various routes towards \textit{inducing} ferromagnetism in modified Hubbard models have been explored, including the introduction of multiple orbitals, nearest-neighbor Coulomb repulsion, longer-range hoppings, or dispersionless (``flat'') bands in the spectrum 
\cite{strack1995exact,tasaki1998nagaoka,vollhardt1999metallic}.

\begin{figure}[tb]
    \centering
    \includegraphics[width=0.7\linewidth]{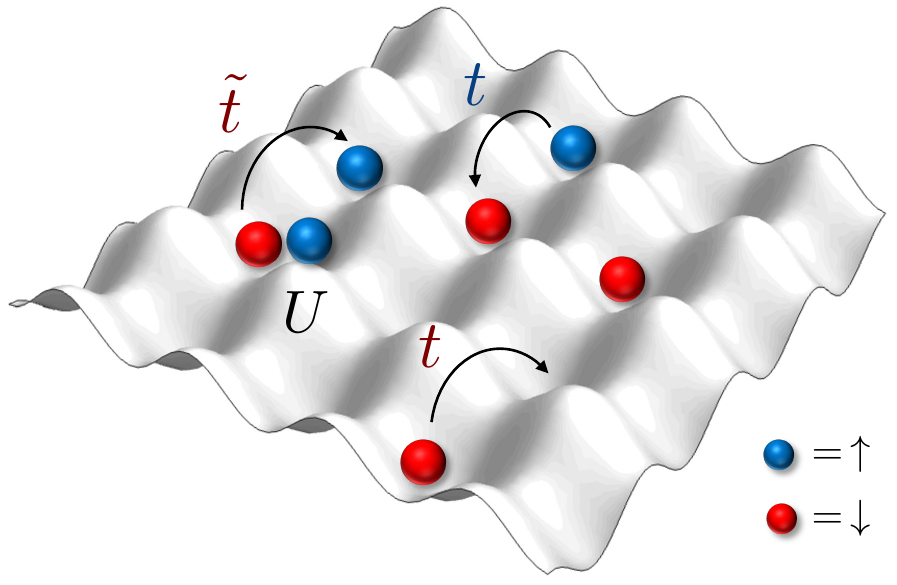}
    \caption{Schematic illustration of the Hubbard model and its extended variant, as described in Eqs.~\eqref{eq:Hmodel} and \eqref{eq:model}, respectively. Fermions hop on an optical lattice with an onsite repulsive interaction $U$ and a tunneling amplitude $t$. We also consider a generalization of this model in which the tunneling matrix element is enhanced to $\tilde{t}$ for a hopping process from a doubly occupied site to an already singly occupied site. In the limit $\tilde{t} = t$, the modified Hamiltonian  reduces to that of the regular Hubbard model.}
    \label{fig:model}
\end{figure}

In this work, we start by investigating the simple Hubbard model---without any of the aforementioned modifications---on a triangular lattice, which can be readily realized in today's optical lattice setups. The Hamiltonian is given by
\begin{equation}
\label{eq:Hmodel}
    H = - t \sum_{\langle i, j\rangle, \sigma} \left(c^\dagger_{i\sigma} c^{\pdagger}_{j \sigma} + \mathrm{h.c.}\right) + U \sum_i n^{\pdagger}_{i\uparrow}n^{\pdagger}_{i\downarrow},
\end{equation}
where $n_{i,\sigma}$\,$\equiv$\,$c^\dagger_{i,\sigma} c^{\pdagger}_{i,\sigma}$ denotes the number operator on site $i$, $n_i $\,$=$\,$n_{i,\uparrow}$\,$+$\,$n_{i,\downarrow}$ is the total occupation of site $i$, and we only consider nearest-neighbor hopping. Studying the phase diagram of this model in the strong-correlation limit $U$\,$\gg$\,$t$, using large-scale density-matrix renormalization group (DMRG)~\cite{PhysRevLett.69.2863,*PhysRevB.48.10345,*RevModPhys.77.259,*schollwock2011density} calculations, we discover extended regions of ferromagnetic order in finite-sized strips. We characterize its intrinsic connection to geometric frustration and discuss its competition with other proximate orders, such as striped phases. Our results thus directly address the recent experiments by~\citet{xu2023frustration} observing ferromagnetism in a frustrated Fermi-Hubbard magnet and account for the origins thereof. Additionally, we demonstrate a route towards inducing ferromagnetism in the square lattice, which is not geometrically frustrated: this can be achieved in a setup where  the second electron on any site of the lattice is much more weakly bound than the first and we examine the prospects for realizing such systems in optical lattices.

\textit{Triangular lattice.---}Our search for elusive ferromagnetism begins with the triangular lattice motivated by Ref.~\onlinecite{mattis1981theory}, which established two conditions for the geometric constraints required to host Nagaoka ferromagnetism: the underlying lattice should have loops, and different paths of particles or holes around such loops should interfere constructively. On the unfrustrated square lattice, the shortest such path is a cluster of four sites, which leads to a hopping amplitude $\sim$\,$t^4$. However, it is the nonbipartite triangular lattice which exhibits the \textit{shortest} possible loop length of three. Consequently, the loop-hopping amplitude is proportional to $t^3$ and changes sign when $t$ does (which is equivalent to exchanging electron and hole doping), so we expect to find ferromagnetism only for doping fractions $\delta$\,$>$\,$0$.

We probe the ground states of the Hamiltonian~\eqref{eq:Hmodel} on two-dimensional cylinders of $N$ sites, with dimensions $L_x$\,$\times$\,$L_y$ (in units of the lattice spacing $a$) and open (periodic) boundary conditions along the $\hat{x}$ ($\hat{y}$) direction ~\cite{stoudenmire2012studying}. Using DMRG with a bond dimension of up to $\chi$\,$=$\,$4800$, we ensure the convergence of our results to a truncation error $< 10^{-7}$ throughout.
Earlier dynamical mean-field theory (DMFT) calculations on the Hubbard model found the region of ferromagnetism in parameter space to be maximized for an electron doping concentration of $\delta$\,$=$\,$0.5$~\cite{merino2006ferromagnetism}. While DMFT is exact only in infinite dimensions, we use this filling fraction as a starting point for the Hubbard model~\eqref{eq:Hmodel} as well (note that this high-density regime is in distinction to Ref.~\onlinecite{morera2022hightemperature}, which illustrated the formation of \textit{local} ferromagnetic polarons around a single doublon).

\begin{figure}[tb]
    \centering        
    \includegraphics[width=\linewidth]{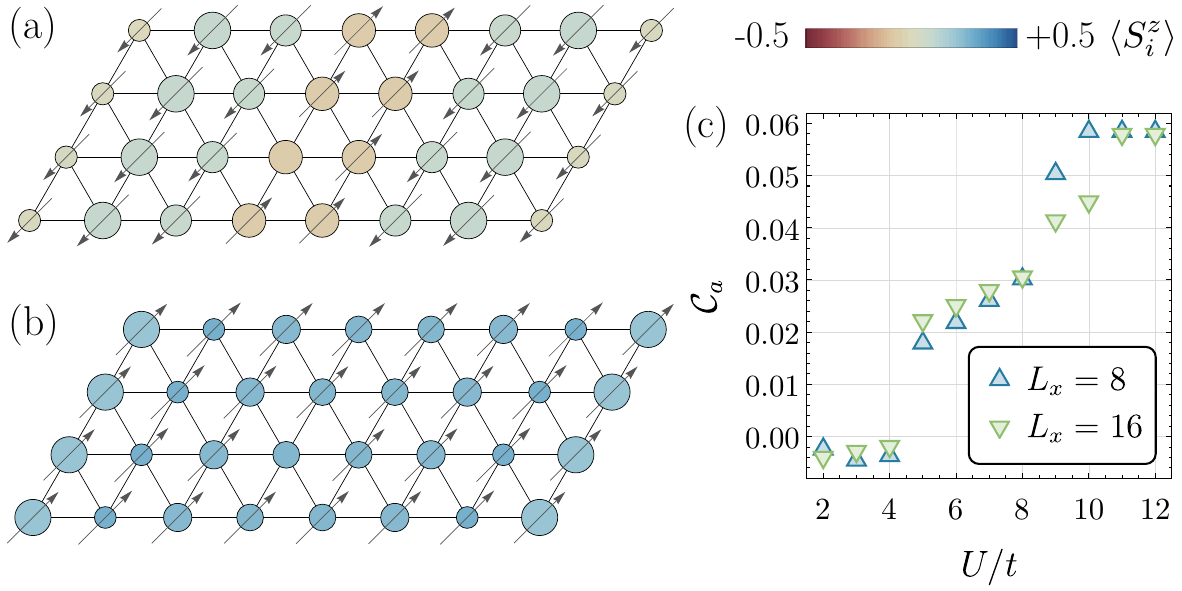}
    \caption{(a) Ground states of the Hubbard model~\eqref{eq:Hmodel} on a triangular lattice for (a) $U/t$\,$=$\,$4$ and (b) $U/t$\,$=$\,$10$, exhibiting striped and ferromagnetic correlations, respectively. (c) The nearest-neighbor correlator $\mathcal{C}_a$, plotted here for two width-$4$ cylinders of different lengths, shows the realization of the Nagaoka state for moderate $U/t$. }
    \label{fig:Hubbard}
\end{figure}

\begin{figure*}[tb]
    \centering        
    \includegraphics[width=\linewidth]{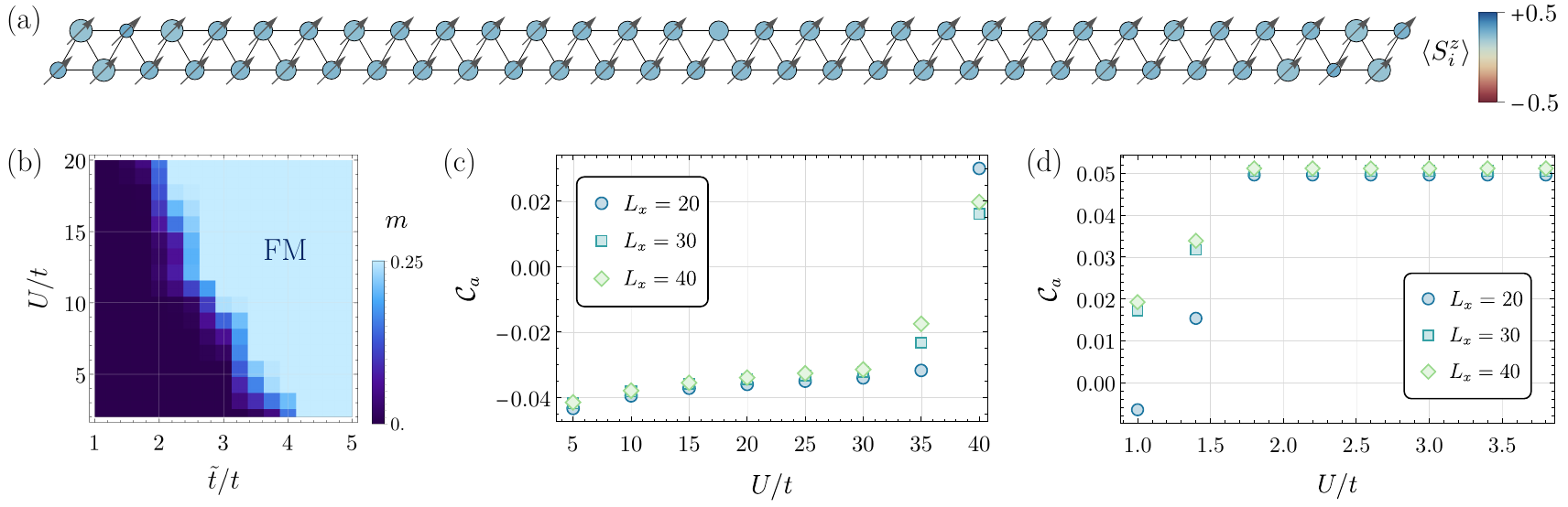}
    \caption{(a) Ground state obtained on a triangular two-leg ladder for $\tilde{t}/t$\,$=$\,$4$, $U$\,$=$\,$0.5 \tilde{t}$ (drawn using the same conventions as in Fig.~\ref{fig:Hubbard}), displaying macroscopic long-ranged ferromagnetic correlations. (b) Phase diagram of a $L_x$\,$=$\,$20$ ladder, as chalked out by the net magnetization, $m$, showing the extent of the Nagaoka state (FM). (c,d) The nearest-neighbor correlator $\mathcal{C}_a$ for ladders of three different lengths, demonstrating the development of ferromagnetism as a function of $U/t$ along the cuts (c) $\tilde{t}/t$\,$=$\,$1$ (the Hubbard limit) and (d) $\tilde{t}/t$\,$=$\,$4$. All calculations presented here are for an electron doping concentration of $\delta$\,$=$\,$0.5$.}
    \label{fig:ladder}
\end{figure*}

Figure~\ref{fig:Hubbard} illustrates the ground states thus obtained  for two representative values of $U/t$, at a doping fraction of $\delta$\,$=$\,$1/2$ above half filling.  Here, the spin operator is defined as $\textbf{S}_i$\,$\equiv$\,$c^{\dagger}_{i\alpha}\vect{\sigma}^{\pdagger}_{\alpha \beta} c^{\pdagger}_{i\beta}$.\,At $U$\,$=$\,$10\, t$, one clearly observes a saturated ferromagnetic ground state, as evidenced by the uniform real-space spin-density profile plotted in Fig.~\ref{fig:Hubbard}(b) (for the sector with the largest $S^z$ quantum number in the degenerate ground-state subspace). 
Interestingly, our calculations uncover a competing ground state with unidirectional charge and spin density modulations, for which the maximal allowed $S^z$ quantum number is zero. An example of this inhomogeneous stripe-ordered state~\cite{zaanen1989charged,machida1989magnetism,kato1990soliton}, which breaks translational and rotational symmetries, is displayed in Fig.~\ref{fig:Hubbard}(a)  for $U$\,$=$\,$4 t$. Similar stripes have also been experimentally observed in certain square-lattice cuprates~\cite{tranquada1995evidence, vojta2009lattice}.
In our case, the origin of the stripes can be understood from the competition between
the domain walls favored by the antiferromagnetic
exchange and the lack thereof preferred for kinetic delocalization. Consider a state with stripes of linear width $\ell$. Relative to the uniform ferromagnet, the increase in the electrons' kinetic energy will be $E_\textsc{k}$\,$\sim$\,$(\delta N)\, \tilde{t}/(\ell L_y)$ while the decrease in interaction energy due to the exchange across the domain walls is $E_\textsc{j}$\,$\sim$\,$-J N/\ell$. The contest between these two energy scales decides between ferromagnetic and stripe orders, with the former eventually prevailing at large enough $U/t$ (as $J$\,$\sim$\,$-t^2/U$).

Having established the existence of the Nagaoka state, we now turn to characterizing its extent and stability. A useful metric to quantify the the ferromagnetic order is the (normalized) average two-point correlation function
$
\mathcal{C}^{}_d =\sum_{\langle i,j \rangle,\, \lvert \lvert \vect{r}_i - \vect{r}_j \rvert \rvert =d}  \langle \textbf{S}^{}_i \cdot  \textbf{S}^{}_j \rangle/\mathcal{N}_b,
$
where the sum runs over $\mathcal{N}_b$ unordered pairs of sites $\langle i, j\rangle$ separated by a distance $d$. In particular, Fig.~\ref{fig:Hubbard}(c) shows the nearest-neighbor correlator $\mathcal{C}_a$ as a function of $U/t$ for two different system sizes. To eliminate the possibility of strong finite-size effects, we consider two different cylinders with one being double the length of the other (results on wider cylinders are documented in the Supplemental Material [SM] \cite{supplement}). In both cases, we see that the nearest-neighbor radial correlator $\mathcal{C}_a$  attains its maximum possible value by $U/t$\,$\simeq$\,$11$ although the onset of ferromagnetic correlations can already be seen for much lower $U/t\sim 6$. We emphasize that  this observed ferromagnetism is truly a many-body phenomenon of the Nagaoka type, as opposed to Stoner ferromagnetism, which occurs whenever the (single-particle) density of states at the Fermi energy $D(E_F)$\,$>$\,$1/U$ and should thus set in for infinitesimal $U/t$\,$>$\,$0$ (as $D(E_F)$  is singular at this filling; see below).

\textit{Two-leg triangular ladders.---}While our results in the previous section manifest the presence of the Nagaoka state, classically, one might suspect this to be an effect of the geometric frustration, which is deleterious to competing antiferromagnetic orders and leads to their suppression. Such a reasoning, however, may break down due to subtle quantum effects; for example, antiferromagnetism can actually be \textit{enhanced} on the triangular lattice by a single hole~\cite{haerter2005kinetic}.
Instead, Nagaoka ferromagnetism on the triangular lattice can be better understood as a consequence of geometrical frustration of the kinetic energy---a strictly quantum-mechanical effect stemming from constructive interference between different paths of a doublon propagating in a spin-polarized background~\cite{haerter2005kinetic}.
A quantitative measure of such kinetic energy frustration is given by $f \equiv W/(2 z \lvert t\rvert)$, where $W$ is the bandwidth and $z$ is the coordination number~\cite{barford1991spinless}, with smaller ratios indicating stronger frustration. The kinetic frustration of even a single triangular cluster results in lowering $f$ below the unfrustrated value of $f$\,$=$\,$1$ for the square lattice~\cite{merino2006ferromagnetism}. For the triangular-lattice Hubbard model, the noninteracting problem has a bandwidth of $9t$ with $z$\,$=$\,$6$, wherefore $f$\,$=$\,$0.75$. The associated $D(E)$ can be computed analytically in terms of a complete elliptic integral of the first kind and exhibits a van Hove singularity at $E$\,$=$\,$-2t$, corresponding to a density of $1.5$ electrons per site. Our earlier choice of $\delta$\,$=$\,$0.5$ thus maximizes $D(E_F)$: a large density of states at the Fermi level lowers the kinetic cost of filling additional single-particle electronic states, thereby enabling the large $U$ (that favors spin alignment) to dominate~\cite{pastor1994electron,pastor1996magnetism,hanisch1997lattice}.

In order to probe the dependence of ferromagnetism on $f$, we now consider a different geometry, namely, a two-leg triangular ladder, which has a reduced coordination number compared to the full lattice. In consistency with the intuition developed above, here, we do not find a ferromagnetic ground state for up to $U/t$\,$=$\,$20$. It is therefore only natural to ask if there are conditions under which ferromagnetism can occur for moderate values of $U/t$ as well.

To answer this question affirmatively, we turn to a generalized Hubbard model---inspired by studies of hydrogenic donors in semiconductors~\cite{nielsen2007nanoscale, nielsen2010search, bhatt1999monte,*berciu2001effects,*bhatt2002diluted}---in which  the second electron on any site of the lattice is much more weakly bound than the first \cite{ghazali1978density,mott1980metal,bhatt1981single}. The Hamiltonian is~\cite{nielsen2007nanoscale,nielsen2010search}
\begin{equation}
\label{eq:model}
    H = -\sum_{\langle i, j\rangle, \sigma} \left( t(n_i,n_j) c^\dagger_{i\sigma} c^{\pdagger}_{j \sigma} + \mathrm{h.c.}\right) + U \sum_i n^{\pdagger}_{i\uparrow}n^{\pdagger}_{i\downarrow};
\end{equation}
the occupation-dependent nearest-neighbor hopping is given by $t(n^{}_i,n^{}_j) = \Tilde{t}$ if $n^{}_i$\,$=$\,$1$,\,$n^{}_j$\,$=$\,$2$ and $t(n^{}_i,n^{}_j) =t$ otherwise. This enhanced hopping (for electron-doped systems) favors the formation of a locally ferromagnetic configuration of spins around the charge, thus yielding a polaronic type of ferromagnetism in the large-$U/t$ limit \cite{samajdar2023polaronic}.

Figure~\ref{fig:ladder}(a) sketches the ground state of this Hamiltonian on a $30$-site-long two-leg ladder for $\tilde{t}/t$\,$=$\,$4$, $U$\,$=$\,$0.5 \tilde{t}$. 
Even for this rather small value of $U/t$, we  unambiguously observe large-scale ferromagnetic order across our entire finite-sized system, characterized by a saturated net magnetization of $m$\,$=$\,$0.25$ with a standard deviation of $0.015$. 
Motivated by this finding, we systematically explore the dependence of ferromagnetism on the Hamiltonian's parameters by constructing a phase diagram based on the magnetization $m$ [Fig.~\ref{fig:ladder}(b)]. We see that increasing $\tilde{t}/t$ results in a dramatic reduction in the minimal $U/t$ needed for the development of ferromagnetic order. We examine the robustness of this behavior by inspecting the two-point correlator $\mathcal{C}_a$ along  two cuts through this phase diagram, for three ladders of varying lengths. For the Hubbard limit $\tilde{t}/t$\,$=$\,$1$ [Fig.~\ref{fig:ladder}(c)], we find that the correlations in the ground state (which is believed to be a Curie-Weiss metal for low $U/t$~\cite{merino2006ferromagnetism}) remain negative till $U/t$\,$\sim$\,$40$ and turn weakly ferromagnetic thereafter. On the contrary, for  $\tilde{t}/t=4$ [Fig.~\ref{fig:ladder}(d)], this transition occurs for a nearly twentyfold smaller $U/t$, irrespective of system size.

Lastly, we note that decreasing (increasing) $\tilde{t}/t$ increases (decreases) the minimum $U/t$ required for a fully polarized ground state. For instance, in stark contrast to Fig.~\ref{fig:ladder}(a), we do not detect any Nagaoka ferromagnet up to $U/t$\,$=$\,$50$  for $\tilde{t}/t$\,$=$\,$2$ (or, as seen earlier, for the Hubbard model at $\tilde{t}/t$\,$=$\,$1$) on the same two-leg cylinders.
This is because ferromagnetism is driven by the relative kinetic energy gain of
a delocalized electron in a background of aligned spins vis-\`a-vis the 
background being in an antiferromagnetic or random configuration,  and this gain is reduced by a smaller hopping amplitude.

\textit{Square lattice.---}Our observation of Nagaoka ferromagnetism on the nonbipartite triangular lattice, which innately lacks electron-hole symmetry, raises the question of whether similar electron-hole asymmetry \textit{induced} by setting $\tilde{t}$\,$\ne$\,$t$ can aid the formation of ferromagnetic states or domains on even (bipartite) square arrays.

At half filling, the ground state of the square-lattice Hubbard model is known to be a Mott insulator with antiferromagnetic N\'eel order for all values of $U/t$\,$>$\,$0$ due to the perfect nesting of the noninteracting Fermi surface. As the doping (either electron or hole) is increased, one expects a transition to a ferromagnetic ground state for sufficiently large $U/t$. Exact diagonalization calculations on small clusters~\cite{nielsen2007nanoscale, nielsen2010search} have previously found that high-spin ground states are obtained at much lower values of $U/t$ for electron doping than for hole doping. Therefore, we consider the electron-doped case hereafter (focusing on the optimal doping fraction of $\delta$\,$=$\,$1/12$ identified in Ref.~\onlinecite{samajdar2023polaronic}) and without loss of generality, set $\tilde{t}/t = 4$.

\begin{figure}[tb]
    \centering
    \includegraphics[width=\linewidth]{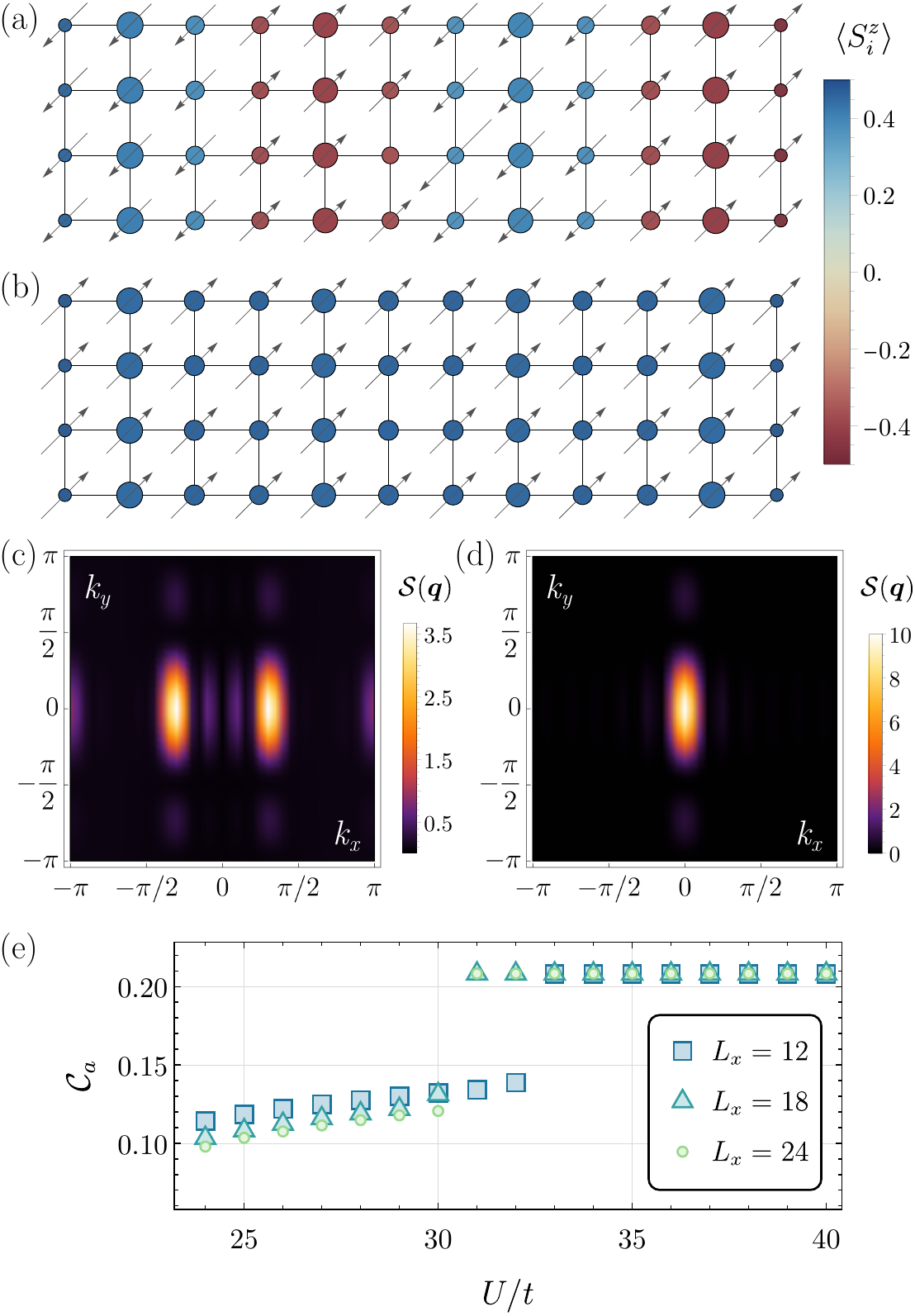}
    \caption{Ground states of the extended Hubbard model \eqref{eq:model} at $\tilde{t}/t$\,$=$\,$4$ for (a) $U$\,$=$\,$6\, \tilde{t}$ and (b) $U$\,$=$\,$9\, \tilde{t}$ on the square lattice. The color of each circle conveys the onsite magnetization $\langle S^z_i\rangle$ while its diameter is proportional to the charge density $\langle n_i \rangle$.  The length of the arrows is proportional to the amplitude of the spin correlation $\langle \textbf{S}_0\cdot \textbf{S}_i\rangle$ with respect to the central site, indexed $0$ (arrow-free). The directions of the arrows encode the sign of $\langle \textbf{S}_0\cdot \textbf{S}_i\rangle$ with arrows pointing northeast (southwest) indicating positive (negative) correlations. (c,d) The static spin structure factors corresponding to the states in (a,b), respectively. (e) The radially averaged nearest-neighbor correlator $\mathcal{C}_a$ at $\tilde{t}/t = 4$ for three width-$4$ cylinders of different lengths, showing the stripe-to-ferromagnet transition. Since the transition occurs between two different magnetically ordered symmetry-breaking states, by Landau theory, we expect it to be first-order, as borne out by the sudden jump in $\mathcal{C}_a$. }
    \label{fig:Square}
\end{figure}

At low $U/t$, we once again find a state with stripelike correlations in the spin density (i.e., with \textit{local} ferromagnetic order), as exemplified by Fig.~\ref{fig:Square}(a) for a $12$\,$\times$\,$4$ cylinder at $U$\,$=$\,$6\,\tilde{t}$. The long-ranged correlations in this state are reflected in the static structure factor
$
\mathcal{S} (\vect{q}) = \sum_{i,j} \langle \textbf{S}^{}_i \cdot  \textbf{S}^{}_j \rangle \exp[i \vect{q}\cdot (\vect{r}_i - \vect{r}_j)]/N,
$
shown in Fig.~\ref{fig:Square}(c). $\mathcal{S} (\vect{q})$ has prominent peaks at $(\pm \pi/3,0)$ indicating a period-$6$ modulation of the spin density along the nonperiodic direction, which accompanies a period-$3$ modulation of the charge density. Across a wide range of $L_x$, we find that the stripes are filled and commensurate with the doping, accommodating exactly one excess electron per unit cell of 12 sites.
Remarkably, at larger $U$\,$=$\,$9\,\tilde{t}$, we see clear evidence of a \textit{global} ferromagnetic ground state, as displayed in Figs.~\ref{fig:Square}(b) and (d) in real and Fourier space, respectively; in this fully polarized state, the structure factor  only exhibits a single peak at $\vect{q}$\,$=$\,$0$ in the Brillouin zone.

To study the stripe--ferromagnet  quantum phase transition, we examine the correlator $\mathcal{C}_a$, which is plotted in Fig.~\ref{fig:Square}(e) as a function of $U/t$ for three different system sizes.
The onset of ferromagnetism is apparent as a sharp increase in $\mathcal{C}_a$ at $U/t$\,$\sim$\,$30$, wherafter it quickly saturates to a system-size-independent value \footnote{The collapse of the data for the two largest system sizes atop one another suggests that our observations are free of finite-size effects along the axial dimension.}. 
Intuitively, the order parameter remains pinned to this maximal value over an extended range of $U/t$ because the gain of band energy due to a single flipped spin does not exceed the concomitant cost from the onsite repulsion; hence, we do not observe a partially polarized ferromagnetic state.

\textit{Discussion and outlook.---}Despite the prediction of Nagaoka ferromagnetism more than half a century ago, its actual realization has proved considerably more challenging. This may partially be attributed to the fragility of the Nagaoka state itself, which was theorized for an infinite system doped with exactly one hole at $U/t$\,$=$\,$\infty$.
In this work, we address a central question: moving away from such an idealized limit, can ferromagnetism occur for finite $U/t$ and macroscopic dopings? Using detailed DMRG computations, we reveal robust ferromagnetism (at least on moderate length scales), analyze its origins, and explore its intimate connections to the underlying lattice geometry. Given that half-filled Hubbard systems are generically either antiferromagnetic (when on a bipartite lattice) or paramagnetic, the emergence of ferromagnetism is indeed a beautiful demonstration of the surprises that strong correlations may engender.

The rarity of Nagaoka ferromagnetism in conventional solid-state materials is well documented. For instance, in many candidate Mott-insulator oxides and chalcogenide systems, $U/t$ is insufficient to generate ferromagnetism. Likewise, strong positional disorder present in doped semiconductors often localizes mobile carriers and inhibits ferromagnetism~\cite{nielsen2010search,bhatt1982scaling}. Optical lattice platforms, on the other hand, offer clean and tunable alternatives to circumvent these issues. In such systems, the lattice depth can be varied to control the tunnelings $t, \tilde{t}$ while the ratio $U/t$ can be  tuned via a Feshbach resonance~\cite{chin2010feshbach}. In the SM \cite{supplement}, we also discuss two possible methods to realize extended Hubbard models akin to Eq.~\eqref{eq:model}, which could pave a new route to obtaining the Nagaoka state.

Even without any such correlated hoppings, our results for the case of $\tilde{t}$\,$=$\,$t$ directly apply to the recent experiments of Ref.~\onlinecite{xu2023frustration}, which probe a square-lattice Hubbard model with a next-nearest-neighbor tunneling of strength $t'$. In the limit $t$\,$\approx$\,$t'$ (when the square lattice is reconstructed to a triangular one),~\citet{xu2023frustration} observe signatures of ferromagnetic correlations for $U/t \approx 9$ with $\delta \approx 0.5$. This is in excellent agreement with our predictions in Fig.~\ref{fig:Hubbard}, thus highlighting the geometric frustration of the kinetic energy as a mechanism for the experimentally observed ferromagnetism.

\begin{acknowledgments}
\textit{Acknowledgments.---}.We thank W.\,S. Bakr, M.\,L. Prichard, U. Schollw\"{o}ck E.\,M. Stoudenmire, and especially J. Kestner for useful discussions. R.S. is supported by the Princeton Quantum Initiative Fellowship. R.N.B. acknowledges support from the UK Foundation at Princeton University and thanks the Aspen Center for Physics, where some of the ideas were conceptualized, for hospitality. The simulations in this paper were performed using the \textsc{ITensor} library~\cite{ITensor} on computational resources managed and supported by Princeton Research Computing, a consortium of groups including the Princeton Institute for Computational Science and Engineering (PICSciE) and the Office of Information Technology's High Performance Computing Center and Visualization Laboratory at Princeton University.

\textit{Note added.---}After the completion of this work, we became aware of two papers also studying Nagaoka ferromagnetism, which recently appeared on the arXiv~\cite{li2023frustration,schlomer2023kinetictomagnetic}.
\end{acknowledgments}

\bibliographystyle{apsrev4-2}
\bibliography{refs.bib}

\newpage
\foreach \x in {1,...,5}
{%
\clearpage
\includepdf[pages={\x}]{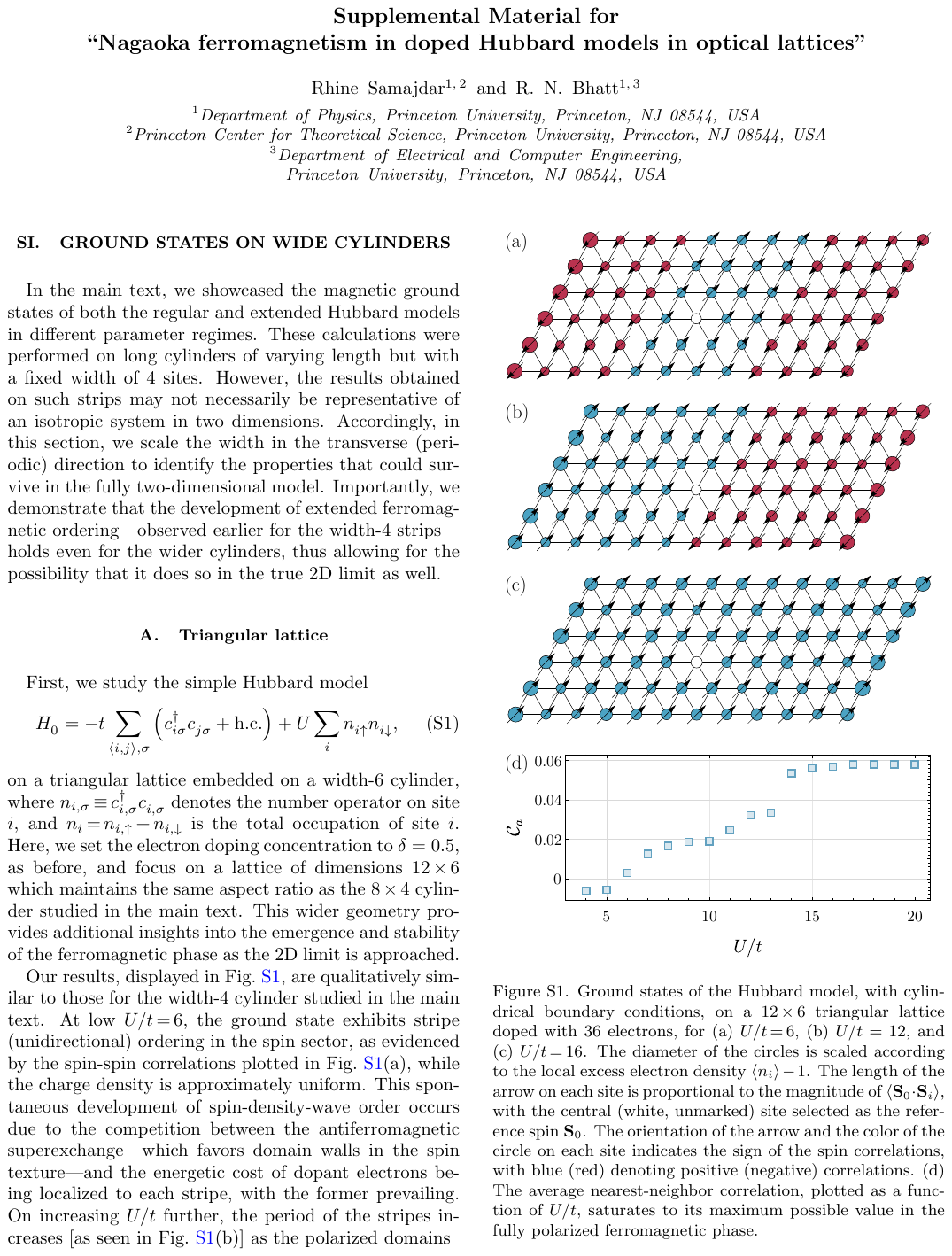} 
}

\end{document}